\title{Non-Restarting CUSUM charts and Control of the False Discovery
  Rate}
\author{Axel Gandy \qquad
  F. Din-Houn Lau\\Department of Mathematics, Imperial College London}
\date{}

\documentclass[11pt]{article}
\usepackage{amsmath,amssymb}
\usepackage{amsfonts}
\usepackage{graphicx}
\usepackage[margin=1in]{geometry}
\usepackage{bm}
\usepackage{natbib}
 \bibpunct{(}{)}{;}{a}{,}{,}
\usepackage{subfig}
\newtheorem{algorithm}{{\itshape Algorithm}}
\newtheorem{thm}{Theorem}

\def\N{\mathbb{N}}
\def\R{\mathbb{R}}
\DeclareMathOperator{\Prob}{pr}

\begin{document}
\maketitle

\begin{abstract}
  Cumulative sum (CUSUM) charts are typically used to detect changes
  in a stream of observations e.g.\ shifts in the mean. Usually, after
  signalling, the chart is restarted by setting it to some value below
  the signalling threshold. We propose a non-restarting CUSUM chart
  which is able to detect periods during which the stream is out of
  control. Further, we advocate an upper boundary to prevent the CUSUM
  chart rising too high, which helps detecting a change back into
  control.  We present a novel algorithm to control the false
  discovery rate (FDR) pointwise in time when considering CUSUM charts
  based on multiple streams of data.  We prove that the FDR is
  controlled under two definitions of a false discovery
  simultaneously. Simulations reveal the difference in FDR control
  when using these two definitions and other desirable definitions of
  a false discovery.
\end{abstract}


\section{Introduction}\label{sec:intro}
One of the most widely used control charts is the cumulative sum
(CUSUM) chart suggested by \cite{page}, which in its simplest form is
defined as follows. Consider observing a stream $X_t$, $t\in
\N=\{1,2,\dots\}$ of independent random variables.  Suppose when in
control $X_t \sim N(0,1)$. Assume that after an unknown time
$\gamma\in [0,\infty]$, the observations switch to an out-of-control
state where $X_t\sim N(\Delta,1)$ for some known $\Delta>0$. Then the
classic CUSUM chart is
\begin{equation}
  \label{eq:classical:cusumchart}
  S_t=\max(S_{t-1}+X_t-\Delta/2,0),\quad S_0=0.
\end{equation}
The chart signals a change at the hitting time $\inf\{t>0; S_t\geq
\zeta\}$ for some threshold $\zeta>0$.
\cite{hawkins1998} give  a  detailed background of CUSUM charts and their applications.

CUSUM charts were originally designed for industrial settings, quoting
\cite{page}:
[Process inspection schemes are] ``required to detect a deterioration
in the quality of the output from a continuous process. When such a
deterioration is suspected some action is taken; for example, the
production may be suspended
and a machine reset.''
This explains why, once a CUSUM chart crosses the threshold $\zeta$, it
is typically restarted at 0. Restarting at a different value
such as $\zeta/2$ has also been suggested
\citep{Lucas}. 

In this paper we are concerned with monitoring multiple data streams
in situations where restarting is not possible, e.g.\ a medical setting
where each stream relates to the performance of a hospital. Even if we
suspect a deterioration of performance, it is unlikely that the
hospital would close or suspend treatment of patients.  Moreover, we
are interested in scenarios where streams  can switch,
potentially multiple times, between an in-control state and an out-of-control
state.  The setting of monitoring
multiple streams of observations has recently become a topic of
increasing interest \citep{Mei2010,Li2009}, in particular in medical
settings \citep{Spiegelhalter2011,Bottle2008Iin,Biswas2008rCi}.

We propose a novel algorithm to control the false discovery rate (FDR)
of multiple data streams pointwise in time. To monitor these data
streams we suggest using non-restarting CUSUM charts with an upper
boundary. A non-restarting CUSUM chart continues when its threshold is
crossed. This leads to periods during which the stream is considered
to be out of control. Moreover, we impose an upper boundary on the
chart which improves detection when the chart comes back in control.

In this algorithm, a false discovery would naturally be defined as
signalling the stream to be out-of-control when in fact the
observations have been in-control since the start. We prove in Theorem
\ref{thm:p_values} that the algorithm simultaneously controls the FDR for the
following less restrictive definition of false discovery:
signalling the stream to be out-of-control when in fact the
observations have been in-control since the last time the chart was at
0.

Previous work concerning FDR control procedures in statistical process
control settings goes back to \cite{BK1} and \cite{BK2}.  \cite{grigg}
considered monitoring normally distributed streams of observations
through CUSUM charts that are restarted after a signal. \cite{Li2009}
propose a method to control the FDR over the stages of a multistage
process. They apply a FDR control procedure on a single unit over the
stages of production with the aim of finding a faulty stage. This
differs from our aim which is the control the FDR pointwise in time
across multiple units. In \cite{Mei2010} a method is proposed using a
global false alarm constraint across multiple streams of
data. However, the setting considered only allows for one global time
at which some of the data streams change from the in-control state to
the out-of-control state.

Our contributions to this area are to focus on a situation where
restarting is not possible, to modify the CUSUM chart to enable it to
signal periods of in-control and out-of-control observations, and to
discuss the meaning of a false discovery in this setting.

\section{Non-Restarting CUSUM Charts with an Upper
  Boundary}\label{sec:CUSUM}
We now present the general setting and CUSUM charts we shall be
using. Consider a stream of independent real-valued random variables
$Z_1,Z_2,\dots$ with distribution functions $F_1,F_2,\dots$
respectively. At time $t$, the random variable, $Z_t$, is in control
if $F_t=F_t^*$ and out of control if $F_t\neq F_t^*$, for some known
in-control distributions $F_1^*, F_2^*,\dots$. We consider extensions
of the CUSUM charts \citep{page} of the form
\begin{equation}
  \label{eq:CUSUM_gen}
  S_t=\varphi\left[\min\left\{\max\left(S_{t-1}+Z_t,0\right),h\right\}\right],\quad S_0=0,
\end{equation}
where $\varphi$ is a non-decreasing function and $h>0$ is a constant
specifying an upper boundary.

The classic CUSUM chart \eqref{eq:classical:cusumchart} reduces to
\eqref{eq:CUSUM_gen} by using $Z_t=X_t-\Delta/2$, with in control
distribution $N(-\Delta/2,1)$, $h=\infty$ and $\varphi(x)=x$.  Another
example is the loglikelihood CUSUM \citep{moustakides} chart
\begin{equation*}
  S_t=\max[S_{t-1}+\log\left\{f_1(X_t)/f_0(X_t)\right\},0],\quad S_0=0,
\end{equation*}
where $f_0$ and $f_1$ are the probability density functions of the
in-control and out-of-control distribution respectively.  Again this
reduces to \eqref{eq:CUSUM_gen} by letting
$Z_t=\log\{f_1(X_t)/f_0(X_t)\}$, $h=\infty$ and $\varphi(x)=x$.

We include $\varphi$ in \eqref{eq:CUSUM_gen} to allow CUSUM charts in
which, at every step, $S_t$ is rounded to finitely many values. For
these charts we can compute the exact distribution of $S_{t}$ at a fixed $t$
using Markov chains \citep{brook}. This is discussed further in Section
\ref{sec:sim}.

We propose not restarting the chart once its threshold is crossed.
Instead, as long as the chart is above the threshold, we say it
signals continuously until it drops back below the threshold. This
will allow us to detect periods where the observations are in or out
of control. To avoid the chart climbing very high above the threshold,
which may make detecting that the stream is back in control difficult,
we impose the upper boundary $h>0$. This is important in our setting
where the observations can switch in and out of control multiple times.

\begin{figure}[tbp]
  \centering
  \includegraphics[width=\linewidth]{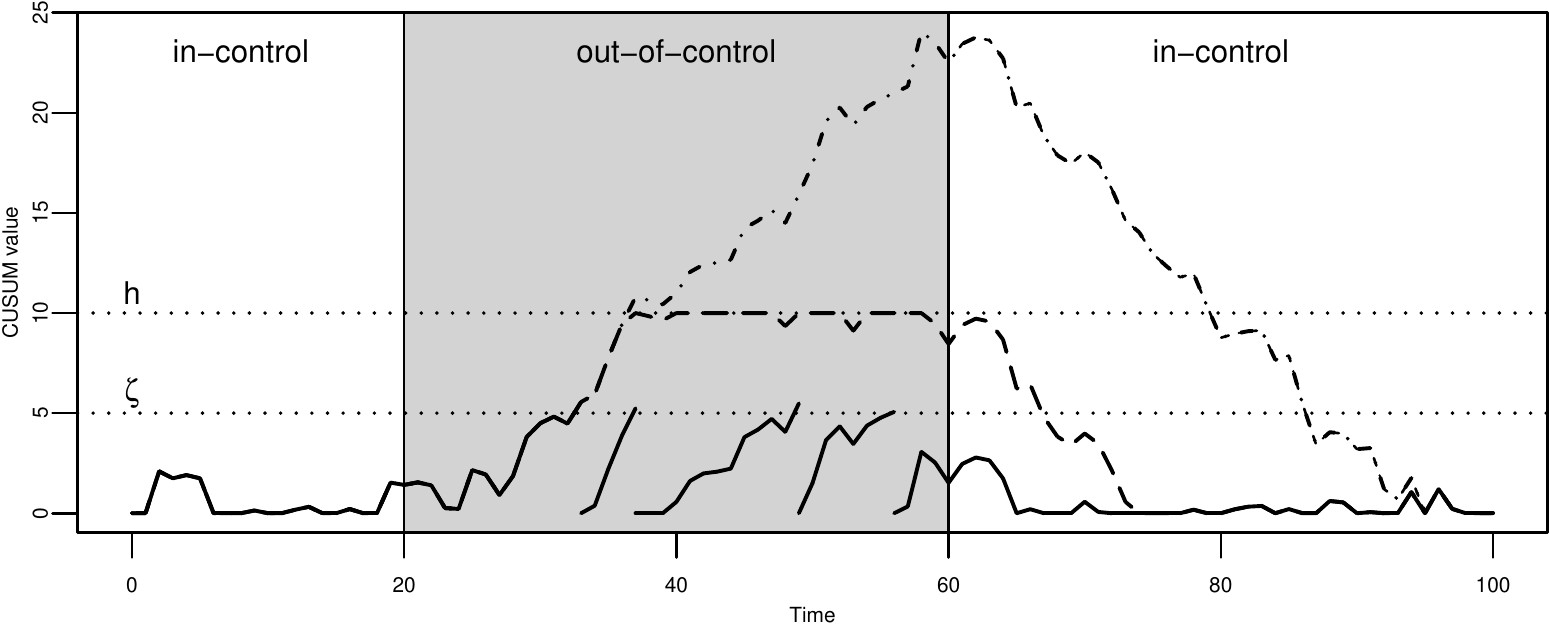}
  \caption{Graph of a CUSUM chart with no upper boundary (dot-dash),
    with upper boundary $h=10$ (dashed) and with a restarting
    threshold $\zeta=h/2=5$ (solid). The grey box represents the times
    at which the observations are truly out-of-control.}
  \label{fig:upper_boundary_CUSUM}
\end{figure}

To compare the non-restarting CUSUM chart to other charts, consider
the CUSUM chart \eqref{eq:CUSUM_gen} with in-control distribution
$N(-1/2,1)$ and out-of-control distribution $N(1/2,1)$ with $h=10$ and
$\varphi(x)=x$. We compare this to the same CUSUM chart with no
upper boundary ($h=\infty$) and a restarting CUSUM chart which resets
to zero when the threshold $\zeta=h/2=5$ is crossed. Figure
\ref{fig:upper_boundary_CUSUM} shows CUSUM charts over 100 time
points, where the observations are out-of-control from time $20$ to
$60$ represented by the grey box.

All charts are identical until they reach the threshold $\zeta$ for
the first time.  The non-restarting chart signals from time 33 to
66. So the out-of-control signal stops a few steps after the stream
has returned to the in-control state.  The restarting chart then
signals at times 33, 37, 49, 56. The main downside of this is that it
does not suggest a period where the stream is out-of-control and,
importantly, there is no signal that the out-of-control period has
ended.  The boundary-free chart signals from 33 to 86. Clearly this
lasts considerably longer than the out-of-control period. This is
mainly due to the high values attained during the out-of-control
period.

\section{False Discovery Rate}
\subsection{Control of False Discovery Rate in Multiple
  Testing}\label{sec:FDR}
We now consider monitoring multiple data streams using a
non-restarting CUSUM chart with upper boundary (Section
\ref{sec:CUSUM}) for each stream. Instead of using a fixed threshold
$\zeta$ to determine which streams are out of control, we suggest
using an FDR control procedure. We first briefly review the procedure
developed by \cite{Benjamini1995}.

Consider testing $N$  null hypotheses
  $H_1^0,H_2^0,\dots,H_N^0$ 
simultaneously.   Denote the
number of true null hypotheses by $m_0$. Let $V$ be the number of true
null hypotheses declared significant and $R$ be the total of null
hypotheses declared significant.  Define $Q=V/R$ as the proportion
of the rejected null hypotheses which are incorrectly rejected, with
the convention $0/0=0$.  The FDR is then defined as $E(Q)$.

Suppose we have $N$ independent tests with corresponding 
$p$-values  $P_1,P_2,\dots,P_N$ for the  hypotheses.
The following algorithm proposed by \cite{Benjamini1995} ensures the
FDR is less than a pre-specified constant $q^*\in(0,1)$.

\begin{algorithm}[Control of the FDR at $q^{*}\in(0,1)$]\label{algo:BH}  \ 
  \begin{enumerate}
  \item Order the $p$-values as $P_{(1)}\leq P_{(2)}\leq \dots \leq
    P_{(N)}$, where $P_{(i)}$ corresponds to $H_{(i)}^{0}$.

  \item Let $k$ be the largest $i$ for which $P_{(i)}\leq \frac{i}{N}q^*$.

  \item Reject $H_{(i)}^0$ for $i=1,2,\dots,k.$
  \end{enumerate}
\end{algorithm}
This procedure controls the FDR at $q^*$ i.e.\ $E(Q)\leq (m_0/N)
q^*\leq q^*$.  The procedure requires \citep[Th.5.1]{FDRdependency}
that the $p$-values satisfy
\begin{equation}
  \label{eq:p-value1}
  \Prob\left(P_i\leq \frac{k}{N}q^*\mid H_i^0\right)\leq \frac{k}{N}q^*\quad (k=0,\dots,N; i=1,2,\dots,N),
\end{equation}
which is satisfied when $P_{i}$ is computed conditionally on
$H_{i}^{0}$ being true \citep[][pg. 64, Lemma 3.3.1]{Lehmann}. The
allocation of which null hypotheses are true can be random, and the
FDR conditional on this allocation will still be controlled.

Based upon the above method, other FDR control procedures have been
developed, e.g.\ the two-step FDR control procedure \citep[Def.
6]{Benjamini2006}, the adaptive linear step-up procedure \citep[Def.
3]{Benjamini2006} and the adaptive step-down procedure
\citep{Gavrilov2009}. These other procedures involve estimating $m_0$,
by $\widehat{m}_0$ say, before applying the \cite{Benjamini1995}
procedure at level $q^* N / \widehat{m}_0$.

\subsection{Algorithm}\label{sec:method}
We wish to control the FDR at each time point using CUSUM charts for
multiple streams.  We first state the algorithm before precisely
defining a false discovery in our setting.

Suppose we observe $N$ independent streams of
observations $(Z_{i,t})_{t\in \N}$ $(i=1,\dots,N)$. Each $Z_{i,t}$ has
distribution function $F_{i,t}$ with $F_{i,t}=F_{i,t}^*$ when
$Z_{i,t}$ is in-control and $F_{i,t}\neq F_{i,t}^*$ when $Z_{i,t}$ is
out-of-control. All $F_{i,t}^*$ are assumed to be known.  For each
stream $(Z_{i,t})_{t\in \N}$ we run a non-restarting CUSUM chart
$S_{i,t}$ with upper boundary $h$ according to \eqref{eq:CUSUM_gen}.

We propose the following algorithm to control the FDR at level
$q^{*}\in(0,1)$ at each time $t$. Any FDR control procedure that
controls the FDR at $q^*$ if \eqref{eq:p-value1} is guaranteed, can be
used. These include the aforementioned two-step, adaptive linear
step-up and adaptive step-down procedures.

The following algorithm is written for the homogeneous
case where  $F_{i,t}^*=F_t^*$  for all $i$.  
\begin{algorithm}[Control of the FDR at
  $q^*\in (0,1)$ at a fixed time $t$]\label{algo:cusum_fdr}
  \ 
  \begin{enumerate}
  \item Let $(S_\nu^{*})_{\nu\in\mathbb{N}}$ be a chart with all
    observations in control, i.e.  $F_{\nu}= F_{\nu}^{*}$ for all $\nu$.
    Compute the distribution of $S_t^{*}$ and let
    $P(s)=\Prob(S_t^{*}\geq s)$.
    
  \item For the observed streams ($i=1,\dots,N$) compute the $p$-values
    $P_{i,t}=P(S_{i,t})$.
    
  \item Apply the chosen FDR procedure with level $q^{*}$ to the $p$-values
    $P_{1,t},\dots P_{N,t}$ . The rejected streams are signalled to be
    out-of-control.
  \end{enumerate}
\end{algorithm}
It is straightforward to adapt this to the general case, where each
stream can have a different in-control distribution or a different
upper boundary, by computing the $p$-values separately for each
stream.

If we use $\varphi$ in \eqref{eq:CUSUM_gen} to force the chart to take
only finitely many values then Step 1 can be accomplished using Markov
chains.  Otherwise, $P(s)$ can be approximated through various methods
such as a finite-state Markov chain approximation \citep{brook} or use
of the steady state distribution of the CUSUM chart \citep{grigg}.

\subsection{Null Hypothesis: In-Control Since Start}\label{sec:null1}
In this section we show that Algorithm \ref{algo:cusum_fdr} in Section
\ref{sec:method} controls the FDR at a fixed time $t$ if a false
discovery is defined as:\ a stream that signals out-of-control at time
$t$, when it has in fact been in control since   time $0$.

To phrase this in the language of hypothesis testing, the null
hypotheses are
\begin{equation}
  \label{eq:null_hypo1}
  H_{i,t}^{0}=\left\{ F_{i,\nu}= F_{i,\nu}^{*}\text{ }\text{ for all } 0<\nu\leq t\right\}
  \quad (i=1,\dots,N).
\end{equation}
A null hypothesis $H^0_{i,t}$ is declared significant when it is
rejected by the FDR control procedure. Thus, at each time
$t\in\N^{0}=\{0,1,2,\dots\}$,
\begin{equation*}
  V\!=\!\#\left\{i\!:\! F_{i,\nu}\!=\! F_{i,\nu}^{*} \text{ for all }
    0<\nu\leq t\text{, } H_{i,t}^{0}\text{ is significant}  \right\}\text{ and } R\!=\!\#\left\{\text{significant hypotheses}\right\}.
\end{equation*}
The $p$-values are computed  in agreement with the null
hypotheses \eqref{eq:null_hypo1}. Thus condition \eqref{eq:p-value1}
holds and our algorithm (Algorithm \ref{algo:cusum_fdr} in Section
\ref{sec:method}) controls the FDR at  $q^{*}$, i.e.  $
E(Q)=E\left(V/R\right)\leq q^{*}$.
 
\subsection{Null Hypothesis: In-Control Since Visiting
  0}\label{sec:null2}
The definition of a false discovery in the previous section implies
that all discoveries made after a stream goes out of control for the
first time are considered true discoveries. Thus a signal for a stream
that has been out-of-control and then comes back in-control will never
be considered a false discovery, no matter how long it has already
been back in control.

In this section we show that Algorithm \ref{algo:cusum_fdr}, without
changing in the way the $p$-values are computed, also controls the FDR
when a false discovery is defined as:\ a stream being signalled
out-of-control at time $t$, when it has been in control since its
chart was at $0$. The corresponding null hypotheses are
\begin{equation*}
  \widetilde{H}_{i,t}^{0}\!=\!\left\{\text{there exists } \tau\!\in\! \left\{0,\dots,t\right\}\text{:\ }
    S_{i,\tau}\!=\!0, F_{i,\nu}\!=\!F_{i,\nu}^{*}\text{ for all }
    \tau<\nu\leq t \right\} \quad (i=1,\dots,N).
\end{equation*}
Thus,
\begin{equation*}
  V\!\!=\!\#\!\left\{i: \widetilde{H}_{i,t}^{0}\text{ is significant and }\text{there exists } \tau\!:
    S_{i,\tau}\!=\!0, F_{i,\nu}\!=\! F_{i,\nu}^{*}\text{ for all }
    \tau<\nu\leq t \right\}\!.
\end{equation*}
The definitions of declared significant and $R$ remain the same as
before. The $p$-values are computed as before. The following theorem
shows that \eqref{eq:p-value1} is satisfied and thus the
\cite{Benjamini1995} FDR procedure still controls the FDR.
\begin{thm}\label{thm:p_values}
  For all $x\in[0,1]$ and for $t\in\N^{0}$,
  \begin{equation*}
    \label{eq:p-value_mycond2}
    \Prob(P_{i,t}\leq x \mid \widetilde{H}_{i,t}^{0})\leq x \quad (i=1,\dots,N).
  \end{equation*}
\end{thm}
\noindent The proof can be found in Appendix 1. To summarize Theorem
\ref{thm:p_values}, the FDR with respect to both sets of hypotheses,
$H_{i,t}^0$ and $\widetilde{H}_{i,t}^0$, is being controlled
simultaneously.

\section{Simulations}\label{sec:sim}
In this section we demonstrate the performance of our proposed method
(Algorithm 2) under different definitions (Section \ref{sec:null1} and
Section \ref{sec:null2}) of a false discovery via simulations.

For each stream, we construct a CUSUM chart according to
\eqref{eq:CUSUM_gen}. In this simulation we let $F_{i,t}^*\sim
N(-1/2,1)$ and $F_{i,t}\sim N(1/2,1)$ when out-of-control, for all
$i,t\in\mathbb{N}$ and set the upper boundary $h=10$.

To compute the in-control CUSUM chart distribution, $S_t^*$, we use
\cite{brook} method. If the chart is forced to take only finitely many
values, by using the function $\varphi$ in \eqref{eq:CUSUM_gen}, then
the distribution can be computed exactly, as it is just the
distribution of a finite-state Markov chain. We proceed by
partitioning $[0,h]$ into the $M+1$ states by using
\begin{equation*}
  \label{eq:varphi}
  \varphi(x)=
  \begin{cases}
    0 & x\in [0,w_1)\\
    (w_j+w_{j-1})/2& x\in [w_{j-1},w_j)\quad(j=2,\dots,M)\\
    h & x\in [w_M,h]
  \end{cases}
\end{equation*}
where $w_j=\frac{h}{M}(j-\frac{1}{2})$ for $j=1,\dots,M$.

For each iteration we took $N=100$ streams over a period of 100 time
points and partitioned $[0,h]$ into 100 states with $q^*=0.05$. A
discrete time-homogeneous Markov chain is used to simulate the
observations, for all charts, moving from in-control to out-of-control
and vice versa. This Markov chain is defined by the transition
probabilities $\Prob(F_{i,t+1}=F_{i,t+1}^*\mid F_{i,t}\neq
F_{i,t}^*)=\alpha$ and $\Prob(F_{i,t+1}\neq F_{i,t+1}^*\mid
F_{i,t}=F_{i,t}^*)=\beta$ for some known $0\leq \alpha,\beta\leq 1$
and for all $t\geq 0$ with all streams starting in control. In this
simulation we let $\alpha=0.01$, $\beta=0.07$. This simulation was
repeated 10,000 times, using the same seed. We consider the
\cite{Benjamini1995}, the two-step and the adaptive linear step-up FDR
control procedures.
\begin{figure}[tbp]
  \centering
  \subfloat[]{\label{fig:thresholds}\includegraphics[width=0.5\textwidth]{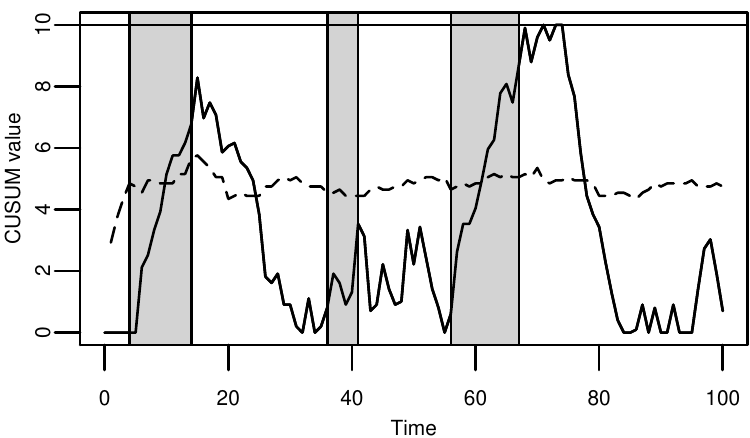}}%
  \subfloat[]{\label{fig:m0s}\includegraphics[width=0.5\textwidth]{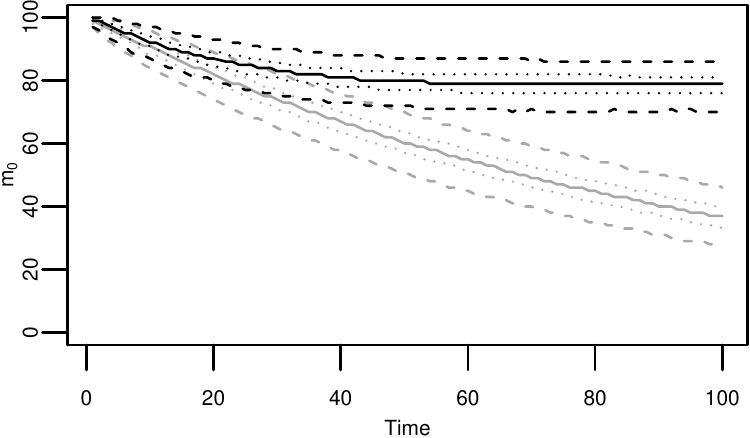}}

  \subfloat[]{\label{fig:control_compare_h0}\includegraphics[width=0.5\textwidth]{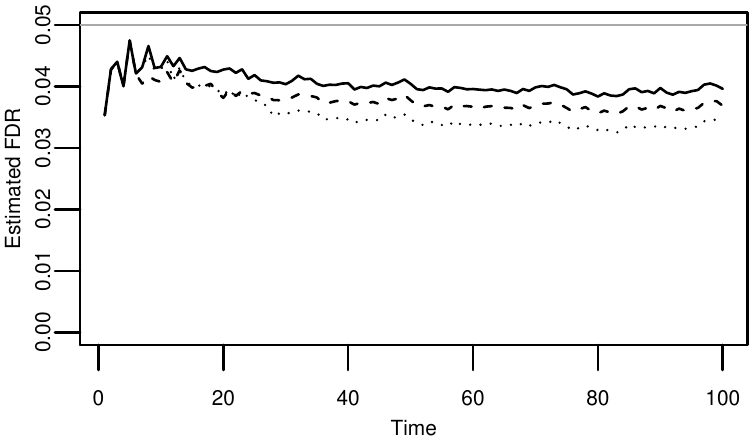}}%
  \subfloat[]{\label{fig:control_compare_h1}\includegraphics[width=0.5\textwidth]{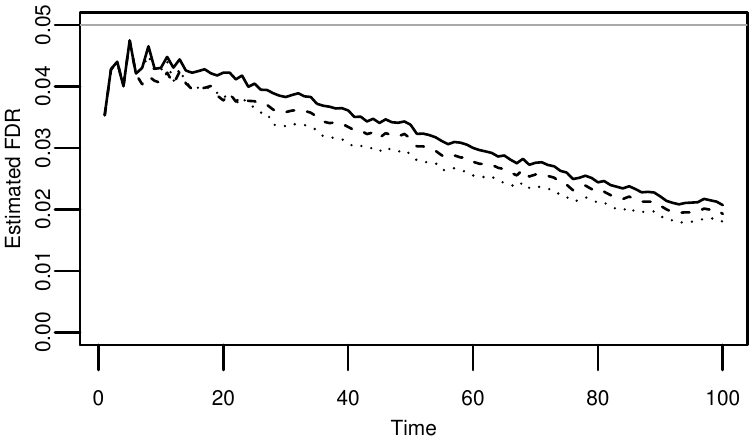}}

  \caption{ (a) Example of a single CUSUM chart (solid) from the simulation with
    thresholds (dashed).  The true out-of-control periods are given by
    the grey areas.  (b) Median of $m_0$ (solid) with 95\% (dashed)
    and 50\% (dotted) quantile pointwise in time under $H_{i,t}^0$
    (grey) and $\widetilde{H}_{i,t}^0$ (black).  (c) 
    Estimated FDR for the \cite{Benjamini1995} (dotted), two-step
    (dashed) and adaptive linear step-up (solid) control
    procedures  with
    $q^*=0.05$ using  $H_{i,t}^0$. (d) same as (c) but using $\widetilde{H}_{i,t}^0$.}
  \label{fig:FDR}
\end{figure}

Figure \ref{fig:thresholds} displays a CUSUM chart from a single
iteration. The threshold given by the \cite{Benjamini1995} FDR control
procedure pointwise in time is also displayed. This threshold, based
upon the remaining 99 charts in the same iteration, is the value which
the presented CUSUM chart needs to exceed in order to signal
out-of-control.

Figure \ref{fig:control_compare_h0} displays the FDR using these
control procedures. All procedures control the FDR below $q^{*}=0.05$.
However, the two-step and the adaptive linear step-up procedures control
the FDR nearer to $q^*$ than the \cite{Benjamini1995} FDR
procedure. This is because other FDR control procedures estimate $m_0$
first, then apply the \cite{Benjamini1995} procedure. For the same
simulation, Figure \ref{fig:control_compare_h1} displays the FDR under
the original hypotheses, $H_{i,t}^0$. We see the FDR for all the
control procedures decreases over time, unlike in Figure
\ref{fig:control_compare_h0}. This is explained by the lower number of
true null hypotheses, $m_0$, at each time point under $H_{i,t}^0$
(Figure \ref{fig:m0s}).

\section{Discussion}
In the simulations in Section \ref{sec:sim}, we have used $\varphi$ to
force the CUSUM chart to take only finitely many states. This ensures
that the distribution of $S_{t}^{\ast}$ in Step 2 of Algorithm
\ref{algo:cusum_fdr} can be computed exactly and thus the FDR is
guaranteed to be controlled.  Allowing the CUSUM chart to take
continuous values, by using $\varphi(x)=x$, will no longer guarantee
the control of the FDR as Step 2 of Algorithm \ref{algo:cusum_fdr} can
only be done approximately. Further simulations, not reported here,
showed that the false discovery rate was still controlled when using a
Markov chain approximation with a reasonably large number of states.
These simulations were similar to those in Section \ref{sec:sim}.

Ideally, we would like to define a false discovery as signalling out
of control at time $t$ when in fact the observation is in control at
time $t$, i.e.\ $F_{i,t}=F_{i,t}^*$.  This is much stronger than our
definitions of a false discovery - and thus the FDR will not be
controlled under this stronger definition. It seems reasonable to assume
that this FDR will depend on how quickly the observations switch between
the in-control and the out-of-control state. Investigating this is a topic 
for further research.

\appendix
\section*{Appendix 1}
\section*{Proof of Theorem 1}\label{ap:proof1}
Since each stream is independent we can drop the subscript $i$.  We
say a random variable $V$ is stochastically smaller than a random
variable $Y$, denoted by $V\leq_{st}Y$, if $\Prob(V\leq x)\geq
\Prob(Y\leq x)$ for all $x\in\R$.

We start the proof by showing, by induction on $t\in \N^0$, that
\begin{equation}
  \label{eq:target2}
  S_t\mid \widetilde{H}_t^{0} \leq_{st} S_t^*.
\end{equation}
At time $t=0$, we have $S_0= S_0^*=0$ and
$\Prob(\widetilde{H}_0^0)=1$, thus \eqref{eq:target2} holds.

At time $t\in\N$ consider the case $F_t\neq F_t^*$.  Then
$\widetilde{H}_t^0=\{S_t=0\}$ and $\Prob(S_t\leq x \mid
\widetilde{H}_t^0)=1$ $\text{ for all } x\in\R$.  Thus
\eqref{eq:target2} holds for this case. For the case $F_t= F_t^*$,
first assume \eqref{eq:target2} holds at time $(t-1)$.  Hence, by the
recursive definition of $S_t$ and $S_t^*$ in (\ref{eq:CUSUM_gen}), and
by the persistence of stochastic orders under convolution of
independent random variables and under action of multiple increasing
functions (Theorems 1.2.13 and 1.2.17 in \citealp[pg.\ 6 and
7]{stoyan}), we get $ S_t\mid
H_{t-1}^0=\varphi\left[\min\{\max(0,S_t+Z_t),h\}\right] \mid
\widetilde{H}_{t-1}^0
\leq_{st} %\varphi\left[\min\{\max(0,S_{t-1}^*+Z_t),h\}\right]=
S_t^*.  $ Thus it suffices to show
\begin{equation}\label{eq:suffice}
  S_t\mid \widetilde{H}_t^0 \leq_{st} S_t\mid
  \widetilde{H}_{t-1}^0.
\end{equation}
As $F_t= F_t^*$, we have $\widetilde{H}_t^0=\widetilde{H}_{t-1}^0\cup
\{S_t=0\}$.  Letting $G(x)=\Prob(S_t\leq x\mid \widetilde{H}_t^0)$,
$J(x)=\Prob(S_t\leq x\mid \widetilde{H}_{t-1}^0)$ and
$\alpha=\Prob(\widetilde{H}_{t-1}^0)/\Prob(\widetilde{H}_t^0)$, we
have
\begin{align}\nonumber
    G(x)&=\Prob(\{S_t\leq x,\widetilde{H}_{t-1}^0\}\cup \left\{S_t=0\right\})/ \Prob(\widetilde{H}_t^0)\\\nonumber
    &=\left\{ \Prob(S_t\leq x,\widetilde{H}_{t-1}^0)+\Prob(S_t=0)-\Prob(S_t=0,\widetilde{H}_{t-1}^0) \right\}\big/\Prob(\widetilde{H}_t^0)\\\nonumber
    &=\left\{
      J(x)\Prob(\widetilde{H}_{t-1}^0)+\Prob(S_t=0)-J(0)\Prob(\widetilde{H}_{t-1}^0)
    \right\}\big/\Prob(\widetilde{H}_t^0) \\\label{eq:cdfs}
    &=\alpha J(x)-\alpha J(0)+\frac{\Prob(S_t=0)}{\Prob(\widetilde{H}_t^0)}.
  \end{align}
By setting $x=0$ in \eqref{eq:cdfs}, we get $
G(0)=\Prob(S_t=0)/\Prob(\widetilde{H}_t^0)$, and so $
G(x)-G(0)=\alpha\left(J(x)-J(0)\right)$.

The distribution of $S_t\mid \widetilde{H}_t^0$ is derived from the
distribution of $S_t\mid \widetilde{H}_{t-1}^0$ by potentially adding
mass at 0 before rescaling. Thus $0<\alpha\leq 1$ and $G(0)\geq J(0)$.
Therefore, $G(x)-G(0)\geq J(x)-J(0).$ Hence, for all $x\in \R$, $
G(x)\geq J(x)+\left\{G(0)-J(0)\right\}\geq J(x).  $ Thus
\eqref{eq:suffice} holds. This finishes showing \eqref{eq:target2}.

Since $P(\cdot)$, defined in Section \ref{sec:method}, is a decreasing
function, application of $P$ on \eqref{eq:target2} (an extension to
Theorem 1.2.13 in \citealp[pg. 6]{stoyan}) yields
\begin{equation}
  \label{eq:so_geq}
  P_t\mid \widetilde{H}_t^0\geq_{st} P_t\mid H_t^0\quad (t\in\N^0).
\end{equation}
By construction of $P_t$, we have
\begin{equation}
  \label{eq:halfway}
  P_t\mid H_t^0\geq_{st} U,
\end{equation}
where $U$ is uniformly distributed on $[0,1]$. Combining
\eqref{eq:so_geq} and \eqref{eq:halfway} gives
\begin{equation*}
  P_t\mid \widetilde{H}_t^0\geq_{st} P_t\mid H_t^0 \geq_{st} U.
\end{equation*}

\bibliography{bibil}
\end{document}